\documentstyle[prl,tighten,aps,epsfig,graphics,multicol,amstex]{revtex}

\draft
\begin{document}

\title{Tripartite entanglement and quantum relative entropy}
\author{E. F. Galv\~{a}o${}^{1}$, M. B. Plenio${}^{2}$ and S. Virmani${}^{2}$}
\address{${}^{1}$ Centre for Quantum Computation, Clarendon Laboratory, Univ. of Oxford, Oxford OX1 3PU,UK}
\address{${}^{2}$ Optics Section, The Blackett Laboratory, Imperial College, London SW7 2BW, UK}
\maketitle
\begin{abstract}
We establish relations between tripartite pure state entanglement
and additivity properties of the bipartite relative entropy of
entanglement. Our results pertain to the asymptotic limit of local
manipulations on a large number of copies of the state. We show that
additivity of the relative entropy would
imply that there are at least two inequivalent types of asymptotic
tripartite entanglement. The methods used include the application
of some useful lemmas that enable us to analytically calculate the
relative entropy for some classes of bipartite states.
\end{abstract}

\draft
\begin{multicols}{2}

\section{Introduction}

In recent years the theory of quantum information and entanglement
processing has developed rapidly.  In the process our
perception of entanglement has changed significantly.
Entanglement used to be regarded just as a surprising
manifestation of the non-locality of quantum mechanics, but today
it is considered as a resource that can be exploited to implement
novel quantum information processing tasks at spatially separated
locations \cite{Plenio V98}. As a resource, entanglement can
appear in many different forms and may not be available in the
specific form necessary for the chosen task. It is therefore
natural to tackle the problem of the interconversion of different
forms of entanglement using local operations and classical
communication only (LOCC). The local concentration of pure
bipartite entanglement has already been considered in the
asymptotic limit, i.e. when large numbers of entangled pairs are
available \cite{Bennett BPS 96}. In this limit it was shown that
any partially entangled state can be \emph{reversibly} converted
into a smaller number of maximally entangled singlet or EPR
states. This remarkable result demonstrates that the entanglement
of any pure bipartite state is essentially equivalent to that of
the singlet state. One can therefore say that the set
$G_{2}=\{|EPR\rangle _{AB}\}$ containing an EPR pair between
systems $A$ and $B$ is  a minimal reversible entanglement
generating set (MREGS) for all bipartite pure states \cite{Bennett
PRST 99}.

It is natural to ask whether there are more inequivalent
forms of entanglement when one considers multi-partite pure state
entanglement in the asymptotic limit; in other words, the problem
is that of identifying an MREGS for multi-partite systems.
Recently it has been shown that indeed GHZ states are inequivalent
to EPR states   in the asymptotic limit  ,
i.e. there is no asymptotically reversible local procedure that
allows the conversion of EPR states into GHZ states \cite{Linden
PSW 99}. Therefore, a MREGS for tripartite systems must contain at
least the GHZ state and the three possible EPR's between any two
of the parties. However, the question as to whether EPR states and
GHZ states form the only kinds of tripartite entanglement, or in
other words whether the set
\begin{equation}
    G_{3} = \{\left|  EPR\right\rangle _{AB},\left|  EPR\right\rangle _{AC},\left|
    EPR\right\rangle _{BC},\left|  GHZ\right\rangle _{ABC}\} \label{G3}
\end{equation}
is an MREGS remained unanswered in
\cite{Bennett PRST 99} and \cite{Linden PSW 99}. The conjecture
that $G_{3}$ as given in eq. (\ref{G3}) forms an MREGS has been
supported by work showing that reversible LOCC on this set yield
Schmidt decomposable states \cite{Bennett PRST 99} and also a
family of states discussed in \cite{Vidal DC 00}. Very recently,
however, Wu and Zhang \cite{Wu Z 00} have shown that without other
effects \cite{neg}, not all four-partite states can be reversibly
built using LOCC on the set of eleven maximally entangled states
of two, three and four parties. Nevertheless, the structure of the
MREGS for tripartite systems remains unknown.

In addition to the developments just described, some relations
have been established \cite{Linden PSW 99} between multipartite
pure state entanglement and a bipartite entanglement measure known
as the \textit{Relative Entropy of Entanglement} \cite{Vedral P
98,Relent,Rains 98}. In this paper we strengthen these relations
further, obtaining new results relating the additivity of the
relative entropy and the structure of the MREGS for tripartite
states.   In section II we summarize the results of \cite{Linden
PSW 99} and  present a number of useful Lemmas that allow us to
exploit symmetries of a quantum state to allow the analytic
computation of the relative entropy of entanglement. In section
\ref{sec G3 mregs} we assume the working hypothesis that the set
$G_{3}$ is an MREGS and derive a series of consequences that would
follow; in particular, we show that the relative entropy of
entanglement (with respect to separable states) would need to be
subadditive for some classes of 2-qubit states. Since to date
there has been no evidence of such subadditivity for qubit states
\cite{Werner}, in section \ref{sec E additive} we adopt the
alternative hypothesis of additivity and explore the consequences,
in particular we discuss implications for the cardinality of the
tripartite MREGS. In section \ref{sec conclusion} we present some
final remarks.

\section{Relative entropy, tripartite entanglement and symmetries}

In this section we introduce some of the notation that we will use
in the remainder of the article.  In the first subsection we summarize the
results of \cite{Linden PSW 99} and in the
second subsection we present some useful
Lemmas that we will employ later on.

\subsection{Basic notation and concepts}
The relative entropy of $\rho$ with respect to any $\sigma$ is
defined as
\begin{equation}
S(\rho\Vert\sigma):=\mbox{tr}(\rho\mbox{log}\rho)-\mbox{tr} \;
(\rho\mbox{log}\sigma)
\end{equation}
 This allows us to define what we mean by the
\textit{Relative Entropy of Entanglement} and \textit{Additivity}:

\smallskip1) \textit{Relative Entropy of Entanglement:}\\
  For bipartite systems this entanglement measure can
take three different forms, E$_{S}$, E$_{PPT}$ or E$_{ND}$
\cite{multirelent}. They are defined as
\begin{displaymath}
    E_{X}(\rho_{AB}):=\min_{\sigma_{AB}\in D(X)}S
    \left(  \rho_{AB}\Vert\sigma_{AB}\right)
\end{displaymath}
where $X=S,PPT,ND$ and the minimum is taken over the set $D$ of
separable(S), Non-Distillable(ND), or Positive Partial
Transpose(PPT) density matrices \cite{Vedral P 98,Relent,Rains
98,Vedral 99}. These measures can further be `regularised' for use
in discussions involving asymptotic manipulations:
\begin{displaymath}
    E_{X}^{reg}(\rho_{AB}):=~\mbox{lim}_{n \rightarrow\infty}(1/n)
    \min_{\sigma_{AB}\in D(X)}S\left(  \rho_{AB}^{\otimes
    n}\Vert\sigma_{AB}\right) \; .
\end{displaymath}
It is important to note that in the case that $\rho$ is either a
pure state or a separable state then all the measures are equal:
$E_{S}(\rho)$ = $E_{PPT}(\rho)$ = $E_{ND}(\rho)$ =
$E_{S}^{reg}(\rho)$ = $E_{PPT}^{reg}(\rho)$ =
$E_{ND}^{reg}(\rho)$.

2) \textit{Additivity:} There are two major types of additivity
which will concern us in this paper:   \\ a) If an
entanglement measure $E$ satisfies $E^{reg}(\rho)$ = $E(\rho)$ we
will say that $E$ is an \textit{asymptotically additive}
measure;  \\ b) If an entanglement measure $E$
satisfies for all $\rho_{1},\rho_{2}$ the relation $E(\rho
_{1}\otimes\rho_{2})=E(\rho_{1})+E(\rho_{2})$ then we say that $E$
is a \textit{fully additive} measure.

The connection between   the relative entropy of
entanglement   and multipartite entanglement was first
pointed out in \cite{Linden PSW 99}, where it was shown that if
two multiparty pure states can be reversibly interconverted then
  the relative entropy of entanglement
must remain constant for any two parties $i,j$. This remarkable
result can be used to derive contraints that must hold if the set
$G_{3}$ is an MREGS for tripartite pure states. In particular,
suppose that we reversibly and asymptotically wish to create a
tripartite pure state $|\Psi_{ABC}\rangle$ between parties A, B
and C, and that per output copy of $|\Psi_{ABC}\rangle$ we will
use $g$ GHZ states and $s_{ij}$ EPR pairs between parties $i$ and
$j$. Then, denoting the reduced density matrices of parties
$i$,$j$ by $\rho_{ij}$, we find that the following relationships
must hold:
\begin{align}
    E_{X}^{reg}(\rho_{ij})  &  =s_{ij}\label{corcond}\\
    S(\rho_{A})  &  =g+s_{AB}+s_{AC}\label{corcond2}\\
    S(\rho_{B})  &  =g+s_{AB}+s_{BC}\\
    S(\rho_{C})  &  =g+s_{AC}+s_{BC}, \label{corcond3}
\end{align}
where $S(\rho_{i})$ represents the Von Neumann entropy of the reduced density
matrix of party $i$ \cite{error}.

It is an open question whether all tripartite states satisfy the
equations (\ref{corcond}-\ref{corcond3}). Any counterexample would
be a state which cannot be generated reversibly from the set
$G_{3}$, representing a new kind of asymptotic tripartite
entanglement. Unfortunately there are no known general techniques
for calculating $E_{X}^{reg}(\rho)$. Despite this , we were able
to obtain some progress in establishing relations between
additivity questions and the structure of the MREGS for tripartite
pure states. In particular, we present classes of states which are
potential candidates for violating relations
(\ref{corcond}-\ref{corcond3}).

\subsection{Symmetries and continuity} \label{sub2}

In this subsection we prove a number of useful Lemmas that
simplify the computation of the relative entropy of entanglement
for states that possess symmetries. In addition, we state a Lemma
due to Donald and Horodecki concerning the continuity of the
relative entropy of entanglement.

We begin by  recalling a Lemma by Rains \cite{Rains
98} which enables us to use local symmetries of the state
$\rho_{AB}$ to narrow down the possible set of optimal states.
Then we extend this Lemma to non-local symmetry
operations.

\textbf{Lemma 1}\cite{Rains 98} If a bipartite density matrix is
invariant under a sub-group of local unitary transformations, then
the optimal PPT state can also be chosen to be invariant under the
same sub-group.

Although the proof can be found in \cite{Rains 98} we present it
here to clarify how this theorem can be generalized to non-local
symmetry groups.

\textbf{Proof} Let there be a bipartite density matrix $\rho$
which is invariant under a sub-group of local transformations G =
$\{ U_{i}\otimes V_{i} \}$ , with an optimal PPT state $\sigma$. For
simplicity, let us assume that the group is discrete (the
generalization to continuous groups is straightforward \cite{Rains
98}). Then $E_{S}(\rho)$ is given by
\begin{equation}
    E_{S}(\rho)=S(\rho||\sigma)=S(\rho||U_{i}\otimes V_{i}\sigma
    U_{i}^{\dagger }\otimes V_{i}^{\dagger}),
\end{equation}
due to the invariance of the relative entropy under unitary
transformations and the invariance of $\rho$ under G. We can then
define another state $\sigma_{s}$ such that
\begin{equation}
    \sigma_{s}={\frac{1}{|G|}}\sum_{i=1}^{|G|}U_{i}\otimes V_{i}\sigma
    U_{i}^{\dagger}\otimes V_{i}^{\dagger}.
\end{equation}

It is important to note that $\sigma_{s}$ is by construction both
PPT and invariant under any unitary transformation selected from G
(due to the rearrangement theorem \cite{tinkham}).

The convexity of the relative entropy \cite{Wehrl 78} hence
implies that

\begin{equation}
    S(\rho||\sigma_{s})\leq{\frac{1}{|G|}}\sum_{i=1}^{|G|}S(\rho||U_{i}\otimes
    V_{i}\sigma U_{i}^{\dagger}\otimes
    V_{i}^{\dagger})=S(\rho||\sigma)
\end{equation}

As $\sigma$ was already an optimal PPT state, this equation must
in fact be a strict equality. Hence $\sigma_{s}$ is also an
optimal PPT state for $\rho$, and by construction is invariant
under the same group G $\square$

\textbf{Corollary 1} This lemma can in fact be extended to include
symmetry groups which include non-local operations and even
operations which are non-physical (such as transposition). Suppose
that there is a set of operations $\{ \Lambda\} $ of the symmetry
group which is either non-local or non-physical, but still takes
density matrices to density matrices. Then the above reasoning
still applies as long as an optimal PPT state $\sigma$ exists for
which $S(\rho|| \Lambda\sigma) = S(\rho|| \sigma)$ and
$\Lambda(\sigma)$ is still PPT.

\smallskip\textbf{Lemma 2} Any disentangled state $\sigma$ which is optimal
for a NPT state $\rho$ must give a partial transposed state
$\sigma^{\Gamma}$ with at least one zero eigenvalue.\newline

\textbf{Proof} Consider a NPT state $\rho$ for which the optimal
PPT state is $\sigma$. Then the convexity of the relative entropy
implies that for all $p~\in~[0,1)$
\begin{align}
S(\rho||p\sigma+(1-p)\rho)  &  \leq p S(\rho||\sigma)+(1-p)S(\rho
||\rho)\nonumber\\ &  < S(\rho||\sigma).
\end{align}

This means that $p\sigma+(1-p)\rho$ must necessarily be a NPT
state $\forall$ $p\in[0,1)$, otherwise $\sigma$ would strictly not
be an optimal PPT state. Now if all the eigenvalues of
$\sigma^{\Gamma}$ are non-zero, then it would be possible to mix
with $\sigma$ a small amount of $\rho$ and still keep the
resulting density matrix PPT. Hence at least one of the
eigenvalues of $\sigma^{\Gamma}$ must be zero $\square$

Finally we state a Lemma concerning the continuity of the relative
entropy of entanglement which is due to Donald and Horodecki
\cite{don}.

\textbf{Lemma 3} $E_{S}(\rho)$ is continuous. Denoting
{tr($|\rho_{1}-\rho _{2}|$)} by $\Delta$ we have the inequality
\begin{equation}
    |E_{S}(\rho_{1})-E_{S}(\rho_{2})| \leq2\mbox{log (dim}(
    {\mathcal{H_{AB}}}) )\Delta- 2 \Delta \mbox{log}(\Delta)+ 4
    \Delta, \label{donald}
\end{equation}
where $\mathcal{H}_{AB}$ is the Hilbert space of $\rho_{i}$. The
proof is given by Donald and Horodecki \cite{don} $\square$

\section{Consequences if the set $G_{3}$ is an MREGS}\label{sec G3 mregs}

In this section we discuss some results which must hold if the set
$G_{3}$ turns out to be an MREGS for tripartite states. We start in subsection \ref{sec asymp}
by proving that $E_{S}$ must be asymptotically subadditive for a class of 2-qubit states if $G_{3}$ is an MREGS. Then in subsection \ref{sec full add} we present implications that set $G_{3}$ being an
MREGS would have for full additivity of entanglement measures, and
in subsection \ref{calcul Ereg} we comment on the possibility of obtaining analytic expressions for
$E_{X}^{reg}$ for some classes of states.

\subsection{Asymptotic subadditivity of $E_{S}\label{sec asymp}$}

\textbf{Theorem 1:} If the set $G_{3}$ is an MREGS for all
tripartite pure states then $E_{S}(\rho)$ must be asymptotically
subadditive for some two qubit states.

In order to prove this theorem we will show that there are states of the form
\begin{equation}
|\psi\rangle=e|000\rangle+f|101\rangle+f|110\rangle\label{psi3terms}%
\end{equation}
for which the non-regularised relative entropies $E_{PPT}$ and $E_{S}$ do not
satisfy modified versions of eqs. (\ref{corcond2}-\ref{corcond3}), with
\begin{equation}
    E_{X}(\rho_{ij})=s_{ij}.
\end{equation}
instead of the regularised relative entropies. This will allow us
to draw the conclusion that for the set $G_{3}$ to be an MREGS for
tripartite pure states we need $E_{S}$ to be asymptotically
subadditive for two-qubit states \cite{Werner}.
  The proof of Theorem 1 uses the Lemmas that have been
proven in subsection \ref{sub2} which help to calculate $E_{S}$
using the symmetries of the quantum state eq.(\ref{psi3terms})
 . We compute the reduced density operators of all
subsystems to find (setting $e$ and $f$ real):
\begin{align}
    \rho_{AB}(e^{2},f^{2})= \rho_{AC}(e^{2},f^{2}) = \left(
    \begin{array}[c]{cccc}%
    e^{2} & 0 & 0 & ef\\
    0 & 0 & 0 & 0\\
    0 & 0 & f^{2} & 0\\
    ef & 0 & 0 & f^{2}
\end{array}
\right) \label{rhoab}\\[0.1cm]
\rho_{BC}(e^{2},f^{2})= \left(
\begin{array}[c]{cccc}
    e^{2} & 0 & 0 & 0\\
    0 & f^{2} & f^{2} & 0\\
    0 & f^{2} & f^{2} & 0\\
    0 & 0 & 0 & 0
\end{array}
\right)
\end{align}

Note that if $e$ and $f$ are non-zero then both these states have
Negative Partial Transpose (NPT), which is equivalent to
inseparability for 2-qubit states \cite{PeresPPT,HoroPPT}. In the
following we will only be discussing 2-qubit states, and hence we
will use the terms separable and PPT interchangeably.

We begin by making the assumption that $E_{S}$ \textit{is}
asymptotically additive  for all 2-qubit states. Then from
(\ref{corcond2}-\ref{corcond3}) and the fact that
$\rho_{AB}=\rho_{AC}$ it follows that a necessary condition for
$|\psi\rangle$ to be obtainable from set $G_{3}$ by reversible
LOCC is given by
\begin{equation}
E_{S}(\rho_{AB})+S(\rho_{AB})=E_{S}(\rho_{BC})+S(\rho_{BC}) \label{necessary}%
\end{equation}
where $S(\rho)$ is the von Neumann entropy of the state $\rho$. Example 1 of
\cite{Vedral P 98} gives us that
\begin{equation}
E_{S}(\rho_{BC})=2(f^{2}-1)\log_{2}(1-f^{2})+(1-2f^{2})\log_{2}(1-2f^{2})
\end{equation}
and by direct computation we find that
\begin{align}
S(\rho_{BC})  &  =-(1-2f^{2})\log_{2}(1-2f^{2})-2f^{2}\log_{2}(2f^{2}),\\
S(\rho_{AB})  &  =-f^{2}\log_{2}f^{2}-(1-f^{2})\log_{2}(1-f^{2})\;.
\end{align}
Combining these equations with eq. (\ref{necessary}), we see that
if $E_{S}$ is asymptotically additive for $\rho_{AB}$ and
$\rho_{BC}$, then $G_{3}$ can only be an MREGS for states of the
form (\ref{psi3terms}) if
\begin{eqnarray}
E_{S}(\rho_{AB})&=& \nonumber\\
&&\hspace*{-8ex}(f^{2}-1)\log_{2}(1-f^{2})-2f^{2}\log_{2}(2f^{2})+f^{2}%
\log_{2}(f^{2}) \label{necnew}%
\end{eqnarray}
In the following we are going to bound $E_{S}(\rho_{AB})$ analytically and
show that eq. (\ref{necnew}) is violated, thereby proving Theorem 1.

  Essentially our task is to constrain the closest
(optimal) PPT state $\sigma_{AB}$ to the density operator
$\rho_{AB}$. We will accomplish this using the Lemmas of
subsection \ref{sub2}. We sequentially apply these lemmas in order
to bound $E_{S}(\rho_{AB})$.  We will use symmetry
arguments first. The elements of $\rho_{AB}$ are real, and
therefore invariant under the act of transposition or complex
conjugation in the computational basis. Transposition has two
properties which allow the application of Corollary 1. The first
is that transposition in a product basis does not change the PPT
properties of a state. The second is that as $\rho_{AB}$ is
symmetric, $-\mbox{tr}\{ \rho\mbox{log}(\sigma)\}$ and hence
$S(\rho|| \sigma)$ are also invariant under transposition of
$\sigma$. Therefore we can utilise the above Corollary and demand
that our optimal PPT state be symmetric, and hence also real.
Furthermore, the state $\rho_{AB}$ is invariant under the local
group
\begin{displaymath}
    \mathcal{X}$ = $\{I,\sigma_{z}\otimes \sigma_{z}\}
\end{displaymath}
and the non-local group
\begin{displaymath}
    {\mathcal{Y}} = \{{I,W}\},
\end{displaymath}
where
$W=(|00\rangle\langle00|+|01\rangle\langle01|-|10\rangle\langle10|+|11\rangle
\langle11|)$. Let us first consider the action of $\mathcal{X}$.
Having fixed the elements to be real, it is a straightforward
calculation to show that the
only density matrices which are invariant under $\mathcal{X}$ must be of the form%

\begin{equation}
\sigma_{AB}=\left(
\begin{array}[c]{cccc}
    x & 0 & 0 & v\\
    0 & y & w & 0\\
    0 & w & z & 0\\
    v & 0 & 0 & u
\end{array}
\right)  . \label{sigABCDEF}
\end{equation}
  Restricting our attention to such states, we can now
consider the action of the non-local symmetry group $\mathcal{Y}$.
Applying the element W of $\mathcal{Y}$ to these states is
equivalent to changing $w$ to $-w$. One can easily see that this
transformation does not change the PPT properties of these states.
Therefore, utilizing the above corollary we can also require that
our optimal state be invariant under changing $w$ to $-w$, in
which case we can set $w=0$.   Now we are in a position
to apply Lemma 2 to $\sigma_{AB}$ of eq. (\ref{sigABCDEF}). Having
already set $w=0$ we now require $\sigma _{AB}^{\Gamma}$ to have
at least one zero eigenvalue. If $x$ or $u$ are zero then $v$ must
be zero to maintain positivity of $\sigma_{AB}$. However, this
cannot be the case if the elements $e^{2}$ and $f^{2}$ of
$\rho_{AB}$ are both non-zero, as then $E_{S}(\rho_{AB})$ becomes
infinite. We will only consider this case, as otherwise the state
in eq. \ref{psi3terms} trivially satisfies eq. (\ref{necessary}).
Therefore we can set $x$,$u$ to be non-zero, and the only way that
$\sigma_{AB}^{\Gamma}$ can have at least one zero eigenvalue and
still be
PPT is if $v=\sqrt{yz}$. Hence the optimal state can be made to take the form

\begin{equation}
\sigma_{AB}=\left(
\begin{array}[c]{cccc}
    x & 0 & 0 & \sqrt{yz}\\
    0 & y & 0 & 0\\
    0 & 0 & z & 0\\
\sqrt{yz} & 0 & 0 & 1-x-y-z
\end{array}
\right)  .
\end{equation}
  To optimize the relative entropy of entanglement with
just the three real parameters left we have to solve the following
partial differential equations

\begin{equation}
    {\frac{\partial}{\partial k}}\left\{  -Tr(\rho_{AB}\log_{2}
    \sigma_{AB})\right\}  =0\quad,\quad k=x,y,z \label{partial}
\end{equation}

In general this can only be done numerically as the resulting
equations are nonlinear in the parameters of $\sigma$. However, as
we are merely looking for an example for which eq. (\ref{necnew})
does not hold, we are free to set convenient values for the
parameters of $\sigma_{AB}$ and then $analytically$ solve
equations linear in the parameters of $\rho_{AB}$. A problem with
this technique is that it does not guarantee that the resulting
$\rho_{AB}$ takes the form of (\ref{rhoab}). In fact, any
two-party subsytem of a pure 2x2x2 state must be rank-2, whereas
generic calculations utilising this technique tend to result in
$\rho_{AB}$ of higher rank. A likely explanation for this is the
measure zero of rank-2 states in the Hilbert space of two qubits.
Nevertheless, it is possible to get extremely close to a state of
the form eq. (\ref{rhoab}) using this technique. We analytically
obtained one particular example for $\sigma$ given by
$x=0.4875473233$, $y=0.1286406856$, $z=0.2953073521$.
Unfortunately, the analytical expressions for the parameters of
the NPT state derived in this way are extremely long, so for
brevity we only write the matrix elements and values in subsequent
calculations to 12
significant digits \cite{email}:

\begin{equation}
\rho_{AB}^{a}=\rho_{AB}(2/3,1/6)+10^{-10}\left(
\begin{array}
[c]{cccc}
0.672 & 0 & 0 & 1.32\\
0 & 0 & 0 & 0\\
0 & 0 & -1.67 & 0\\
1.32 & 0 & 0 & 0.995\nonumber
\end{array}
\right)  ,\nonumber
\end{equation}
where $\rho_{AB}(2/3,1/6)$ is defined via eq(\ref{rhoab}). This results in a
relative entropy
\begin{equation}
E_{S}(\rho_{AB}^{a})=.354761489848
\end{equation}
Note that the state $\rho_{AB}^{a}$ is very close to the state $\rho
_{AB}(2/3,1/6)$. Using Lemma 3 one can hence bound the entanglement of
$\rho_{AB}(2/3,1/6)$ as

\begin{equation}
    E_{S}[\rho_{AB}(2/3,1/6)]=.354761489848\pm3.1\cdot10^{-8}.
\end{equation}

This value is not compatible with the prediction of $E=.3167$,
obtained from eq. (\ref{necnew}). This means that the value of the
non-regularised entanglement $E_{S}(\rho_{AB})$ is too high to
satisfy eq.(\ref{necnew}), and hence the set $G_{3}$ can only be
an MREGS for states of three qubits if $E_{S}$ is asymptotically
subadditive, thus finishing the proof of Theorem 1 ${}_{\Box}$

The result above showed that provided $G_{3}$ is an MREGS, then
$\rho_{AB}$ given by (\ref{psi3terms}) with $e^{2}=2/3$ and
$f^{2}=1/6$ must be asymptotically subadditive. It is clear from
the method we used that similar proofs can be made for different
values of $e$ and $f$. We have written a program that calculates
$E_{S}$ using a gradient search algorithm \cite{Vedral P 98} to
test this hypothesis. The results we obtained suggest that, given
the assumption   that $G_{3}$ is an MREGS, $\rho_{AB}$
must be asymptotically subadditive for generic values of $e$ and
$f$. Equations (\ref{corcond}-\ref{corcond3}) are automatically
satisfied when $e=f$ because of the symmetry under permutations of
the three parties.  Using our program we have numerically checked
the additivity of $E_S$ for two copies of the state $\rho_{AB}$
and have found that, within the limits of numerical precision,
additivity is satisfied in this case. This provides some weak
evidence that indeed $E_S$ is additive for this class of states
and that therefore $G_3$ is not an MREGS. The rigorous proof of
this result has, however, not been completed so far.

\subsection{Full additivity of $E_{S}^{reg}$}\label{sec full add}

Although we have shown that $E_{S}$ must be asymptotically subadditive for
$G_{3}$ to be an MREGS, Wu and Zhang recently made a stronger claim that full
additivity of $E_{S}$ is required \cite{Wu Z 00}. Their original discussion, however,
was based on equations (23) of \cite{Linden PSW 99}, and hence also relies on
the implicit assumption of asymptotic subadditivity present in \cite{Linden
PSW 99}. Here we present a corrected version of their result:

\textbf{Theorem 2} If the set $G_{3}$ is an MREGS for tripartite pure states
then $E_{X}^{reg}$ must be fully additive.

\textbf{Proof} The proof is as given in \cite{Wu Z 00}, except replacing
$E_{X}$ with $E_{X}^{reg}$ $\square$

Thus the question of $G_{3}$ being an MREGS can be related to both the
asymptotic subadditivity of $E_{S}$ and the full additivity of $E_{X}^{reg}$.
In fact even stronger statements can be made about the relationship between
the regularised forms of the relative entropy of entanglement:

\textbf{Theorem 3} If $G_{3}$ is an MREGS then $E_{S}^{reg}$=$E_{PPT}^{reg}%
$=$E_{ND}^{reg}$.

\textbf{Proof}  The proof relies on the fact that for pure states
and separable states all of these measures are identical
\cite{Rains 98,Vedral P 98,multirelent,Vedral 99} - they are equal to
the entropy of the reduced density matrix in the case of bipartite
pure states, and are zero for all separable states. Therefore
the right hand sides of conditions (\ref{corcond}
-\ref{corcond3}) are in fact the same for all of these measures.
Hence the left hand sides of these equations must also be the
same, and all regularised versions are equivalent if the set
$G_{3}$ is an MREGS $\square$

Theorem 3 has interesting consequences. Although theorem 1 above
gives an example of a 2x2x2 dimensional (three qubit) state which
cannot be obtained by reversible LOCC on set $G_{3}$ if $E_{S}$ is
asymptotically additive, theorem 3 can in fact be used to give
higher dimensional states with the same property. All PPT bound
\cite{HoroBound} entangled states (which only exist in dimensions
of at least 2-qutrits) have $E_{PPT}^{reg}=0$ and $E_{S} \neq 0$.
Hence any purifications of bound entangled states would only be
reversibly obtainable from $G_{3}$ if $E_{S}$ is asymptotically
subadditive for all bound entangled states.

\subsection{Calculating $E_{X}^{reg}$}\label{calcul Ereg}

If it is true that $G_{3}$ is an MREGS for tripartite pure states,
then we can use relations (\ref{corcond}-\ref{corcond3}) to obtain
$E_{X}^{reg}$ for some bipartite states. A simple example is
obtained from the tripartite state
\begin{equation}
    |\Lambda(a,b)\rangle = a|000\rangle + b|100\rangle + b|101\rangle + b|110\rangle + b|111\rangle ,
\label{lambdastate}
\end{equation}
whose reduced density matrix for parties A and B is
\begin{equation}
    \rho_{AB}= \rho_{AC} = \left(
    \begin{array}[c]{cccc}%
    a^{2} & 0 & ab & ab\\
    0 & 0 & 0 & 0\\
    ab & 0 & 2b^{2} & 2b^{2}\\
    ab & 0 & 2b^{2} & 2b^{2}
\end{array}
\right) .
\label{rhoablambda}
\end{equation}
It is easy to show that $\rho_{BC}$ is separable (it is PPT). Since $\rho_{AB}=\rho_{AC}$, the relations
\ref{corcond}-\ref{corcond3} mean that if $G_{3}$ is an MREGS, then it must be true that
\begin{equation}
E_{S}^{reg}(\rho_{AB})=S(\rho_{BC})-S(\rho_{AB}).
\label{ESlambda}
\end{equation}
A closed formula for $E_{S}^{reg}(\rho_{AB})$ can then be obtained by directly computing the von Neumann
entropies from eqs. (\ref{lambdastate}-\ref{ESlambda}).
Similar calculations can be used to find $E_{X}^{reg}$
for many states, provided they are obtainable from $G_{3}$ by LOCC.

\section{Consequences if $E_{s}$ is asymptotically additive}\label{sec E additive}

In section \ref{sec G3 mregs} we discussed various consequences
that would follow if set $G_{3}$ can be proven to be an MREGS for
tripartite pure states. In particular, we have shown that this
would entail subadditivity of the relative entropy of entanglement
$E_{S}$. Despite extensive numerical work \cite{Vedral P
98,Horodecki STT 99}, no indication of subadditivity of $E_{s}$
has been found to date  for 2-qubit states \cite{Werner} . In this
section we analyze what would follow if $E_{s}$ turns out to be
asymptotically additive for 2-qubit states.

The first and most obvious conclusion would be that the set
$G_{3}$ could not be an MREGS as follows directly from Theorem 1
for 3-qubit states and Theorem 3 for higher dimensional states. It
is interesting to note that the states discussed in Theorem 1 are
examples of the `W'-states of \cite{Dur VC 00}, which have been
shown to be inequivalent to GHZ states in the sense that one class
of states cannot be converted to the other with non-zero
probability (for single copy manipulations). If $E_{S}$ is
asymptotically additive, it follows from Theorem 1 that a similar
inequivalence persists in the asymptotic and reversible case.

There are also implications for the minimum size which the MREGS must have.
We have performed numerical tests which indicate that the tripartite states

\begin{equation}
    |\Lambda(a,b)\rangle = a|000\rangle + b|100\rangle + b|101\rangle + b|110\rangle + b|111\rangle
\end{equation}
do not satisfy equations (\ref{corcond}-\ref{corcond3}) with the non-regularised
relative entropies (w.r.t. separable states). Note that for this state $\rho_{BC}$ can
easily be shown to be separable (it is invariant under partial
transposition), whereas the states of AB and AC are inseparable -
hence party A is the `odd one out'. This implies that should the
relative entropies for the reduced density matrices of this state
be additive, we would require that at least one more state be
added to the set $G_{3}$ to turn it into an MREGS. Moreover, this
state would have to be separable across parties B and C. However,
if one such state were added, due to symmetry we would in fact
require a further two states to be added as well - corresponding
to making the other parties B or C  the `odd ones out' instead of
party A. This would mean that the minimum cardinality of a
tripartite MREGS would have to be seven - the GHZ, three EPR pairs
and the three kinds of the above state  \cite{detail}. A similar argument cannot
be applied for the state of Theorem 1 as it is not separable
between any two parties.

\section{Conclusion}\label{sec conclusion}

Our results strengthen the connection between the bipartite relative entropy of entanglement
 and the structure of the MREGS for tripartite pure
states. In particular, we have shown that in the case that $E_{S}$
turns out to be asymptotically additive  for 2-qubit states , then
symmetry arguments can be used to show that the states we have
investigated cannot be
 obtained reversibly from the set
$G_{3}=\{\left|  EPR\right\rangle _{AB},\left|  EPR\right\rangle _{AC},\left|
EPR\right\rangle _{BC},\left|  GHZ\right\rangle _{ABC}\}$. On the other hand, if $E_{reg}^{X}$ is {\it not} fully additive, then $G_{3}$ also cannot be an MREGS. It would be
interesting to further investigate the relationship between bipartite mixed
state entanglement and pure tripartite entanglement, as results in one area
may bear fruit in the other.

\section{Acknowledgments}

We would like to thank Jens Eisert, Leah Henderson and Lucien
Hardy for valuable discussions. We also thank Matthew Donald for
comments on $E_{S}^{reg}$, and are especially grateful to Max
Sacchi for pointing out some serious mistakes in earlier drafts of
this work. We would also like to thank Reinhard Werner for communicating the results of \cite{Werner} prior to publication. We acknowledge support from the U.K. O.R.S. Awards
scheme, the Brazilian agency Coordena\c{c}\~{a}o de
Aperfei\c{c}oamento de Pessoal de N\'{i}vel Superior (CAPES), the U.K.
Engineering and Physical Sciences Research Council (EPSRC), the
Leverhulme Trust and the European Union project EQUIP.

\end{multicols}

\begin{thebibliography}{99}
\bibitem {error} In \cite{Linden PSW 99} relationships similar to
(\ref{corcond}) were derived using $E_{S}(\rho_{ij})$ in place of $E_{S}%
^{reg}(\rho_{ij})$. Unfortunately this requires the implicit assumption that
$E_{S}(\rho_{ij})$ is asymptotically additive, a result which has not
yet been established.
\bibitem {detail} It might be argued that the GHZ state may be replaced by some other tripartite state which together with EPR pairs could then be used to generate GHZs and the $|\Lambda\rangle$ states. However, in order to be able to generate the GHZ in the case that $E_{S}$ is asymptotically additive, any such state would also have to be separable across any two parties, in which case it could not be used to generate the $|\Lambda\rangle$ states.
\end{thebibliography}
\end{document}